\documentclass[10pt, conference, a4paper]{IEEEtran}
\usepackage[table]{xcolor}
\usepackage{multirow}
\usepackage{orcidlink}
\usepackage{booktabs}
\usepackage{cite}
\usepackage{amsmath,amssymb,amsfonts}
\usepackage[nolist,nohyperlinks]{acronym}
\usepackage[inline]{enumitem}
\usepackage[detect-weight=true, detect-family=true]{siunitx}
\usepackage{algpseudocode}
\usepackage{algorithm}
\usepackage{svg}
\usepackage{multirow}
\usepackage{balance}
\usepackage{graphicx}
\usepackage[nameinlink, capitalise]{cleveref}
\usepackage[usestackEOL]{stackengine}
\usepackage{stfloats}
\DeclareSIUnit\ohm{\ensuremath\Omega}
\DeclareSIUnit\db{dB}
\DeclareSIUnit{\belmilliwatt}{Bm}
\DeclareSIUnit{\bel}{B}
\DeclareSIUnit{\bitpersecond}{bps}
\DeclareSIUnit{\samplepersecond}{Sps}
\DeclareSIUnit{\nothing}{\relax}
\usepackage[percent]{overpic}
\usepackage[export]{adjustbox}

\usepackage{pifont}

\DeclareSIUnit\ohm{\ensuremath\Omega}
\DeclareSIUnit\db{dB}
\DeclareSIUnit{\belmilliwatt}{Bm}
\DeclareSIUnit{\bel}{B}
\DeclareSIUnit{\bitpersecond}{bps}
\DeclareSIUnit{\samplepersecond}{Sps}
\DeclareSIUnit{\nothing}{\relax}

\usepackage{pifont}

\makeatletter
\providecommand*{\AC@hyperref}[2]{#2}
\makeatother

\makeatletter
\newcommand{\linebreakand}{%
  \end{@IEEEauthorhalign}
  \hfill\mbox{}\par
  \mbox{}\hfill\begin{@IEEEauthorhalign}
}
\makeatother

\makeatletter
\hypersetup{
    draft=true
}
\renewcommand{\href}[2]{#2}
\renewcommand{\url}[1]{\nolinkurl{#1}}
\makeatother

\begin{acronym}
    \acro{ADC}{Analog to Digital Converter}
    \acro{BLE}{Bluetooth Low Energy}
    \acro{dToF}{direct Time-of-Flight}
    \acro{FoV}{field of view}
    \acro{GNSS}{Global Navigation Satellite System}
    \acro{GPIO}{General-Purpose Input/Output}
    \acro{HVAC}{heating, ventilation and air conditioning}
    \acro{LDO}{Low-Dropout Regulator}
    \acro{LiDAR}{Light Detection and Ranging}
    \acro{MAE}{mean absolute error}
    \acro{MCU}{Microcontroller}
    \acro{PIR}{passive infrared}
    \acro{RMSE}{root mean square error}
    \acro{SiP}{System in Package}
    \acro{SoC}{System on Chip}
    \acro{SPAD}{single-photon avalanche diode}
    \acro{ToF}{Time-of-Flight}
    \acro{UWB}{Ultra-Wideband}
    \acro{VCSEL}{vertical-cavity surface-emitting laser}
\end{acronym}

\usepackage{fancyhdr}
\fancypagestyle{firstpage}{     
  \fancyhf{}
  \chead{\textcolor{gray}{This work has been submitted to the IEEE for possible publication.}}
}

\begin{document}
\title{PepDeTech: Anonymous Occupancy Monitoring on a Battery-Powered 2.3k-Zone dToF LiDAR Node with On-Chip Processing and Bluetooth LE Reporting\\}

\author{\IEEEauthorblockN{1\textsuperscript{st} Victor Luder}
\IEEEauthorblockA{\textit{PBL-DITET} \\
\textit{ETH Zurich}\\
Zurich, Switzerland \\
victor.luder@pbl.ee.ethz.ch}
\vspace{-1.0cm}
\and
\IEEEauthorblockN{2\textsuperscript{nd} Félicien Ndozi}
\IEEEauthorblockA{\textit{PBL-DITET} \\
\textit{ETH Zurich}\\
Zurich, Switzerland \\
felicien.ndozi@pbl.ee.ethz.ch}
\vspace{-1.0cm}
\and
\IEEEauthorblockN{3\textsuperscript{rd} Michele Magno}
\IEEEauthorblockA{\textit{PBL-DITET} \\
\textit{ETH Zurich}\\
Zurich, Switzerland \\
michele.magno@pbl.ee.ethz.ch}
\vspace{-1.0cm}
}
\maketitle

\thispagestyle{firstpage}
\begin{abstract}
As populations grow, society is increasingly required to shift towards systems of shared resources, which allow space, energy and time to be optimised. Facilitating this transition depends on data: how many people are present, when and where. From such information, schedules can be organised more efficiently and resources allocated where they matter. At the same time, privacy concerns and energy consumption limit deployment: cameras are frequently unacceptable, while simpler sensors miss people and cannot provide the full picture.
This paper presents \textit{PepDeTech}, a compact, battery-powered occupancy node that combines a $54\times42$-zone \ac{dToF} \ac{LiDAR} (\textsc{VL53L9CX}) with a \ac{BLE} \ac{SoC}. Three deterministic algorithms, background subtraction, motion history and a track lifecycle, run on-chip and count the people simultaneously, providing real-time occupancy monitoring. No trained model is involved, and setup takes less than a minute. Over three scenarios, background subtraction reaches a per-frame $\pm1$-person accuracy of \qty{83.6}{\percent} (\ac{MAE} 0.70~persons), tested with up to five people present. At a \qty{30}{\second} interval the node draws only \qty{3.1}{\milli\watt}, lasting \qty{60}{\day} on a \qty{1200}{\milli\ampere\hour} cell. Several weeks of deployment on a single charge make the node usable as a survey instrument for facility management: a room can be characterised for a measurement campaign without cabling, network provisioning or a camera, and the resulting occupancy record is what ventilation, lighting, cleaning and space-allocation decisions are argued from, ultimately resulting in more efficient resource allocation.
\end{abstract}

\acresetall

\vspace{0.0 cm}
\section{Introduction}
Optimising resource allocation in facility management increasingly calls for decisions justified by data rather than by assumption: when to light, heat and cool a space, and how many desks, chairs and meeting rooms are needed. Studies show that lighting and \ac{HVAC} dominate the energy use of commercial buildings and are usually operated on schedules or for an assumed maximum occupancy~\cite{Sun_2020, Lu_2021}. Feeding the actual number of occupants into the control loop has saved \qty{25}{\percent} of \ac{HVAC} energy in a deployment~\cite{Beltran_2013} and up to \qty{45}{\percent} in simulation~\cite{Lu_2021}. Occupancy has been sensed with \ac{PIR}, CO\textsubscript{2}, radio signals~\cite{Depatla_2015}, cameras, thermal arrays and depth sensors~\cite{Sun_2020}, each with a trade-off: \ac{PIR} sensors are responsive only to motion~\cite{Raykov_2016}, environmental sensors react within minutes, and cameras raise privacy concerns~\cite{Lu_2021}. Thermal arrays and depth sensors avoid identifiable images, but the reported systems rely on models trained on labelled data, which reach at most \qty{82.7}{\percent} balanced accuracy~\cite{Xie_2023}, or are mains-powered, mounted at doorways and processed on a single-board computer or server~\cite{Munir_2017, Lu_2021, Jeong_2025}. Low-power applications have combined acceleration data with a single-zone \ac{ToF} sensor~\cite{Luder_2025}, which leaves a gap between counting the exact number of people and keeping energy consumption low.

This paper presents \textit{PepDeTech}, which addresses this gap by integrating a just-released, low-power, high-resolution \ac{dToF} \ac{LiDAR} module, whose depth image allows people to be segmented directly, onto a power-constrained platform. The \textsc{VL53L9CX} resolves the scene into $54\times42$ zones, each covering \qty{1}{\degree}~\cite{ST_VL53L9CX}. While ranging, the sensor draws up to \qty{800}{\milli\watt}~\cite{ST_UM3683}, and still \qty{160}{\milli\watt} in our configuration, which is two orders of magnitude more than the \ac{MCU} power consumption. A battery-powered node is therefore only feasible if the active sensor time is optimised through its settings and exposure time, and all processing is kept on-chip.

\textit{PepDeTech} is a battery-powered, \ac{BLE}-connected occupancy node that counts the people present in its field of view on-chip and broadcasts only the count. It needs no infrastructure and no training data. The node can be placed exactly where a question arises, run there for a few weeks and then be removed, providing the data behind facility-management decisions for exactly the area of interest.

In summary, this paper presents:
\begin{itemize}
\item A battery-powered node combining a $54\times42$-zone \ac{dToF} \ac{LiDAR} and a \ac{BLE} \ac{SoC}, with a switchable sensor power domain.
\item A lightweight on-chip pipeline with three training-free algorithms that share one front end and are evaluated on identical frames.
\item A low-power deployment mode that samples bursts of frames and broadcasts an 8-bit count advertisement.
\item An evaluation over three scenarios, reaching an exact-count accuracy of \qty{53.0}{\percent} and a $\pm1$-person accuracy of \qty{83.6}{\percent}, together with a measured energy characterisation of each algorithm. Background subtraction draws \qty{3.1}{\milli\watt} at a \qty{30}{\second} interval, and the history-based algorithms \num{4.1} times as much. On an \qty{85}{\milli\ampere\hour} cell, this lasts for 4.2 days.
\end{itemize}

\begin{figure*}[!b]
    \vspace{-0.4cm}
    \centering
    \includegraphics[width=0.92\textwidth]{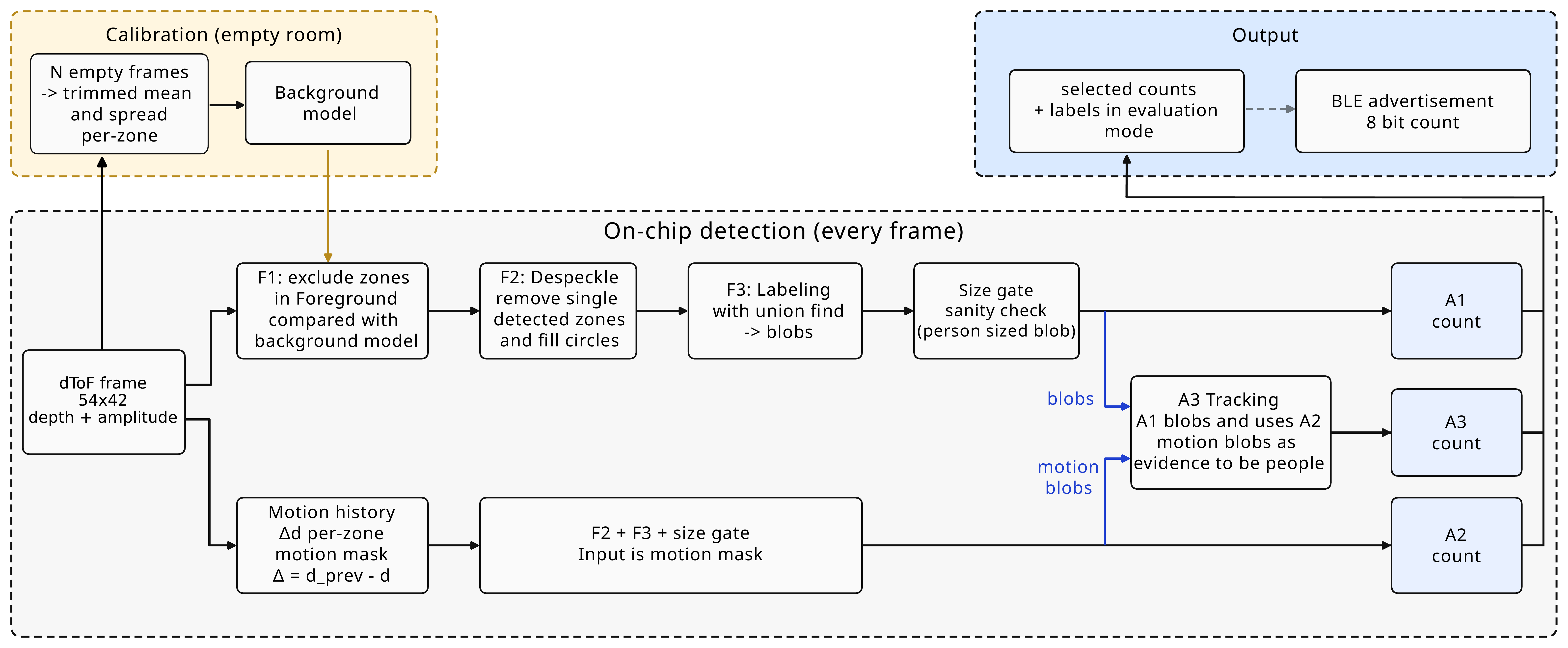}
    \vspace{-3mm}
    \caption{On-chip detection pipeline. The empty-room calibration provides the background model. Each frame passes the shared front end (F1-F3 and size gate), from which A1 counts blobs, A2 segments a motion-history mask, A3 tracks A1's blobs using A2's motion as evidence.}
    \label{fig:pipeline}
\end{figure*}
\section{Related Work}\label{related_work}

\begin{table*}[t]
\centering
\caption{Comparison of state-of-the-art privacy-preserving people counting systems.}
\label{tab:rel-works}
\vspace{-0.3cm}
\renewcommand{\arraystretch}{1.15}
\footnotesize
\resizebox{\textwidth}{!}{
\begin{tabular}{@{}p{2.1cm}p{2.7cm}p{2.7cm}p{2.7cm}p{2.7cm}p{2.7cm}p{3.1cm}@{}}
\toprule
 & \textsc{Mohammadmoradi et al.}~\cite{Mohammadmoradi_2017} &
 \textsc{Huang et al.}~\cite{Huang_2021} &
 \textsc{Lu et al.}~\cite{Lu_2021} &
 \textsc{Xie et al.}~\cite{Xie_2023} &
 \textsc{Jeong et al.}~\cite{Jeong_2025} &
 \textbf{This work} \\
\midrule
Year & 2017 & 2021 & 2021 & 2023 & 2025 & 2026 \\
Sensor &
Thermal array, $8\times8$ &
mmWave radar &
8 × 8 \ac{ToF} sensors &
Thermal array, $8\times8$ &
\ac{ToF} camera &
\ac{dToF} \ac{LiDAR}, $54\times42$ \\
Counting &
Doorway entries/exits &
Doorway entries/exits &
Zone-boundary crossings &
People per frame &
Doorway entries/exits &
People present in the \ac{FoV} \\
Method &
Thresholding, tracking &
Static clutter removal algorithm &
Background subtraction, clustering &
Deep learning, trained on labelled data &
Mean-shift clustering &
Training-free, calibrated per room \\
Processing &
Raspberry Pi Zero &
Raspberry Pi 4 &
Raspberry Pi Zero + central PC &
\ac{MCU} (STM32L4) &
Not reported &
On-chip, \ac{BLE} \ac{SoC} \\
Power/Energy &
Not reported &
Not reported &
Mains &
\qtyrange{17}{120}{\micro\joule}/inference &
Not reported &
Battery, \qty{3.1}{\milli\watt} \\
Node output &
Not reported &
Count &
Raw depth frames &
Not reported &
Not reported &
8-bit count \\
Accuracy &
\qty{93}{\percent} &
\qty{98}{\percent} one person, \qty{65}{\percent} for five &
\qty{0.4}{\percent} error  &
balanced accuracy \qtyrange{55.7}{82.7}{\percent} &
$>$\qty{90}{\percent} &
\qty{53.0}{\percent} exact count accuracy
and within $\pm1$ person accuracy of \qty{83.6}{\percent} \\
\bottomrule
\end{tabular}
}
\vspace{-0.4cm}
\end{table*}

Low-resolution thermal arrays are the most common privacy-preserving imaging sensor for occupancy monitoring. \textit{ThermoSense} has reached a \ac{RMSE} of 0.35~persons by fusing an $8\times8$ array with a \ac{PIR} sensor~\cite{Beltran_2013}. Doorway crossings have been evaluated in many scenarios; Mohammadmoradi et al.~\cite{Mohammadmoradi_2017} have achieved an accuracy of \qty{93}{\percent}. On the other hand, Mikkilineni et al. have built wireless occupancy monitoring nodes that use long-wave infrared (LWIR) focal-plane arrays (FPAs), which detect thermal energy and with it humans ~\cite{Mikkilineni_2019}. Xie et al. ~\cite{Xie_2023} presented a deep-learning approach on infrared arrays, that runs on an STM32L4 at \qtyrange{17}{120}{\micro\joule} per inference. These approaches all use distance and ambient temperature, and some depend on learned models, which work reliably but are restricted to the environments their training data represent.

Depth separates people from the background by geometry, which even single-pixel \ac{ToF} arrays exploit for privacy-preserving tracking~\cite{Bhattacharya_2016}. Heads and shoulders have been detected in Kinect frames on an ARM board at over \qty{99}{\percent} accuracy~\cite{Munir_2017}, zone-boundary crossings counted with sparse $8\times8$ \ac{ToF} sensors at about \qty{0.4}{\percent} error, each read by a Raspberry Pi Zero and processed centrally~\cite{Lu_2021}, and doorway passages counted with a label-free \ac{ToF} camera at over \qty{90}{\percent}~\cite{Jeong_2025}. While mmWave radar tracks people with \qty{98}{\percent} accuracy for one person, it falls to \qty{65}{\percent} for five~\cite{Huang_2021}.

As \cref{tab:rel-works} shows, the depth-based systems count crossings, which may accumulate errors; most run on single-board computers or servers, and do not report node power. \textit{PepDeTech} counts the people present from every measurement, runs entirely on a battery-powered \ac{BLE} \ac{SoC}, and replaces training with a per-room calibration.

\section{System Overview}\label{system_overview}
The \textsc{VL53L9CX} \ac{dToF} \ac{LiDAR}~\cite{ST_VL53L9CX} illuminates the scene with a \qty{940}{\nano\meter} \ac{VCSEL} and reports a distance, a return amplitude and an ambient level per zone over a $\qty{54}{\degree}\times\qty{42}{\degree}$ field of view~\cite{ST_UM3683}.

\begin{figure}[t]
    \centering
    \includegraphics[width=0.92\linewidth]{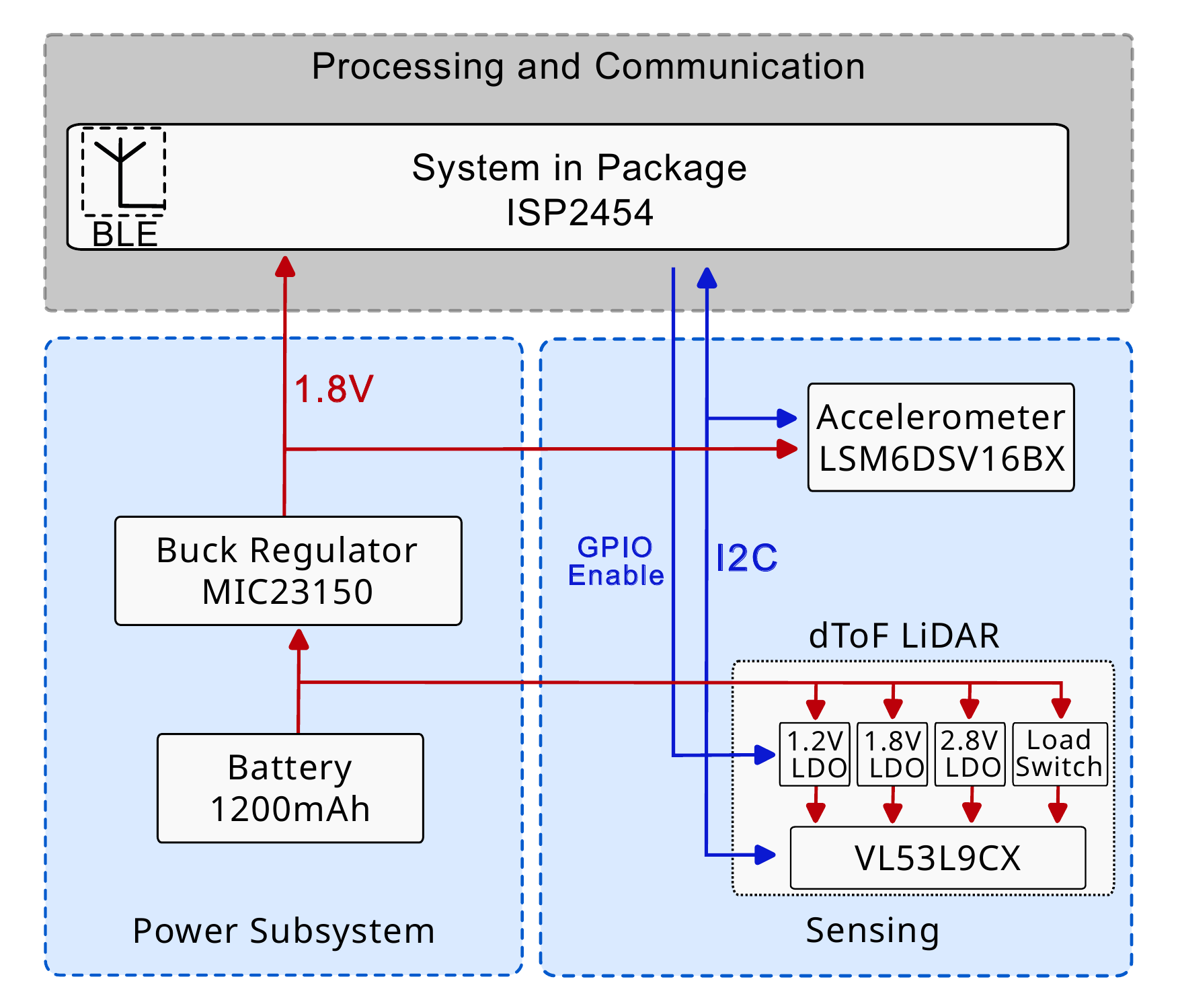}
    \vspace{-3mm}
    \caption{The sensor's three \acp{LDO} and its laser-driver load switch share one \ac{MCU}-controlled enable. Supplies in red, data and control in blue.}
    \label{fig:hardware_setup}
    \vspace{-0.45cm}
\end{figure}

The node is built around the \textsc{ISP2454} \ac{SiP}~\cite{InsightSiP_ISP2454}, which integrates the nRF54L15 \ac{SoC}~\cite{Nordic_nRF54L15} with antenna, crystals and DC/DC converter in an \qtyproduct{8 x 8 x 1}{\milli\meter} package. The sensor is connected to it over I\textsuperscript{2}C at \qty{400}{\kilo\hertz}. As shown in~\cref{fig:hardware_setup}, a \qty{3.7}{\volt} Li-ion battery feeds a buck converter whose \qty{1.8}{\volt} rail continuously supplies the \ac{MCU} and an accelerometer that can be used for motion wake-up. The sensor is supplied separately on the \ac{LiDAR} add-on board: the three \acp{LDO} for its \qty{1.2}{\volt}, \qty{1.8}{\volt} and \qty{2.8}{\volt} rails, together with a load switch passing the cell voltage to the laser driver, share a single \ac{GPIO} enable. The sensor can therefore be disconnected entirely, which is what the duty-cycled mode of~\cref{subsec:power} does.

\begin{figure}
    \centering
    \includegraphics[width=0.92\linewidth]{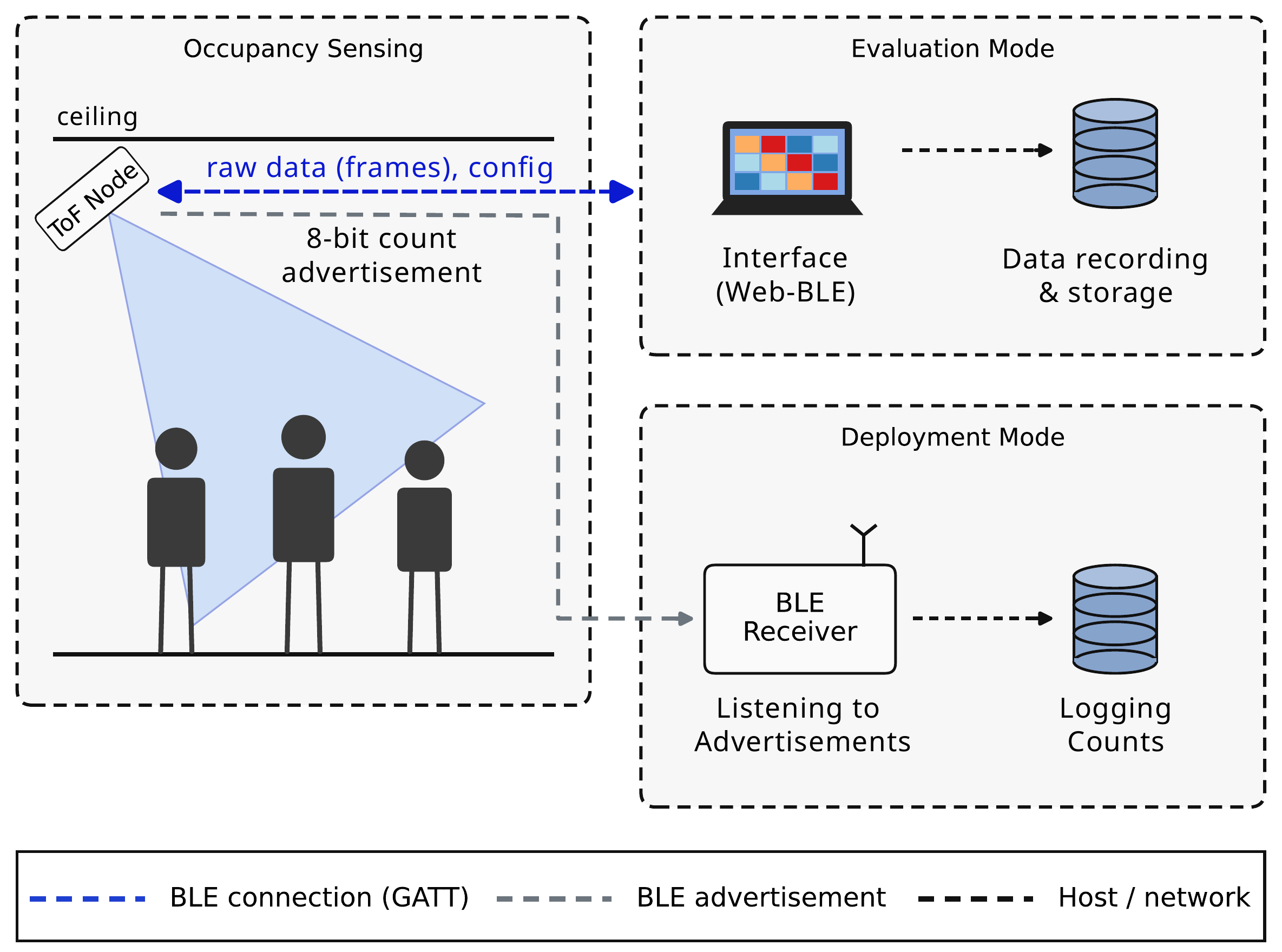}
    \vspace{-3mm}
    \caption{The node counts people on-chip. In evaluation mode a \ac{BLE} connection carries frames, labels and counts to a web interface, which allows recording, visualization and configuration. In deployment mode a passive receiver logs the 8-bit count advertisement.}
    \label{fig:system_overview}
    \vspace{-0.45cm}
\end{figure}

As illustrated in~\cref{fig:system_overview}, the node is mounted in an elevated position. The node has two operating modes, which differ in how \ac{BLE} is used. In evaluation mode, the node streams raw data frames, labels and counts, and accepts configuration parameters, so that the system can be configured and tuned. Once a suitable setting has been found, the node is switched to deployment mode for the lowest power consumption. It then drops the connection and carries only the 8-bit count in the advertisement data, which a passive receiver logs without an active link.

\section{On-Chip Detection Pipeline}\label{sec:algorithms}
The pipeline in~\cref{fig:pipeline} consists of a one-time calibration, a shared front end and three algorithms (A1, A2 and A3) that run on every frame. All stages are deterministic and lightweight: no model is trained, and no labelled data is required. The few parameters, such as the foreground threshold, the size-gate bounds and the motion-history length, have a direct physical meaning and were set once by hand, and all three rooms in~\cref{tab:accuracy} use the same values. The same firmware therefore runs in every room, and only the background model is acquired on site.

\subsection{Empty-Room Calibration}
Over a minimum of $K=16$ frames of the empty room, the node keeps for each zone $i$ the sum, minimum, maximum and number $n_i$ of valid distances $d_i^{(k)}$, and computes the trimmed-mean background $b_i$ and the spread $s_i$:
\begin{equation}
\label{eq:calibration}
\begin{aligned}
b_i &= \frac{\sum_k d_i^{(k)} - \min_k d_i^{(k)} - \max_k d_i^{(k)}}{n_i - 2}, \\
s_i &= \max_k d_i^{(k)} - \min_k d_i^{(k)}.
\end{aligned}
\end{equation}
Zones with $s_i > \qty{300}{\milli\meter}$ (windows, curtains, screens) are \emph{excluded}. The model occupies \qty{11.6}{\kilo\byte} of memory.

\subsection{Shared Front End}
\textbf{F1 Foreground.} In this step, each zone is compared with the background model to detect foreground zones. A zone is tested against the threshold $\theta_i = \max(T_{\min}, s_i)$, with $T_{\min}=\qty{150}{\milli\meter}$. This test is extended by an amplitude check, which classifies each zone into one of three classes: \emph{NEAR}, if the zone is now clearly closer than in the empty room; \emph{APPEARED}, if the zone was invalid in the empty room but now returns a valid distance; and \emph{SHADOW}, if a previously valid zone becomes invalid, for instance because dark clothing absorbs the \qty{940}{\nano\meter} light. A zone that falls into any of these classes is marked as foreground and passed to the next stage.

\textbf{F2 Despeckle.} Each foreground zone is compared with its eight
neighbours. Single detected zones, i.e.\ zones with fewer than two foreground
neighbours, are removed as noise. Conversely, a background zone enclosed by
seven or more foreground neighbours is set to foreground, which fills
single-zone holes inside a blob.

\textbf{F3 Labelling.} 
This step turns the foreground zones into separate objects (blobs), so each person can be counted.
This is done with a two-pass labelling with union-find~\cite{Wu_2009}. First, 4-connectivity detects zones that share the same edge (not diagonal touching) and groups them into the same foreground blob. With this, diagonally touching people stay separate and zones that have the same edge become one blob.

\textbf{Size gate.} Since a zone's footprint grows with distance, each blob $j$ is assigned its real area $A_j = \sum_{i \in j} (p\, d_i)^2$ with $p = \qty{17.45}{\milli\radian}$, and accepted if $\qty{0.08}{\meter\squared} \leq A_j \leq \qty{4}{\meter\squared}$. The same bounds thus reject noise and open doors at any distance.

\subsection{A1 Background Subtraction}
A1 is a classical background subtraction algorithm~\cite{Piccardi_2004}, the approach Lu \textit{et al.} also applied to low-resolution \ac{ToF} occupancy monitoring~\cite{Lu_2021}. Unlike adaptive mixture models that relearn the background continuously~\cite{Stauffer_1999}, the model here is acquired once, which keeps the pipeline deterministic and stops a person who sits still from being absorbed into it. A1 detects static and moving people alike, but an object moved after calibration, such as  a chair, might be counted until recalibration.

\subsection{A2 Motion History}
A2 needs no background. It stores the previous frame and compares the zones. For zones valid in both frames, the signed difference $\Delta_i = d_i^{\,\mathrm{prev}} - d_i$ updates a motion history $H_i$. The construction is the motion history image of Bobick et al.~\cite{Bobick_2001}, introduced for intensity silhouettes and here driven by range instead, which removes its dependence on appearance and lighting:
\begin{equation}
H_i \leftarrow \begin{cases} H_{\max}, & \Delta_i > \tau_i,\\ \max(H_i - 1,\, 0), & \text{otherwise,}\end{cases}
\label{eq:mhi}
\end{equation}
with $H_{\max}=8$ frames. Using only decreasing distances avoids counting the position a person left as a second blob, and holding triggered zones fills the interior of a moving person. The threshold $\tau_i$ is adapted online from non-triggering frames. The motion mask $H_i > 1$ is passed on to F2 and F3. A2 ignores moved furniture but loses people who stop moving.

\subsection{A3 Fusion and Track Lifecycle}
A3 doesn’t search for people itself. It takes A1’s blobs as candidate people, assigns each a track ID and follows it across frames. This tracking-by-detection pattern as in  \textsc{Sort}~\cite{Bewley_2016} was applied by Munir \textit{et al.}~\cite{Munir_2017} for depth-camera occupancy monitoring. A3 tracks the A1 blobs over successive frames and uses the A2 motion blobs as
evidence: an A1 blob that overlaps an A2 motion blob is confirmed as a person, which separates occupants from static objects such as a chair moved after calibration.

\section{Experimental Results}\label{sec:evaluation}

\subsection{Experimental Setup}
The node was mounted facing the room from above at $h=\qty{1.90}{\meter}$ in three scenarios (meeting room, meeting area and office space). The sensor ran at full resolution with \qty{20}{\milli\second} exposure time. To shorten the experiments, accuracy was recorded in free-running mode. Because A1 compares each frame only with the fixed background model, its per-frame accuracy is the same as in duty-cycled deployment. A2 and A3, which depend on frame history, are evaluated here under continuous operation. After calibration in each room (S1--S3 in~\cref{tab:accuracy}), up to five participants were present in the field of view while an observer entered the true count. All algorithms ran on every frame, so all results stem from identical data and parameters.

\begin{figure}[t]
    \centering
    \includegraphics[width=0.6\linewidth]{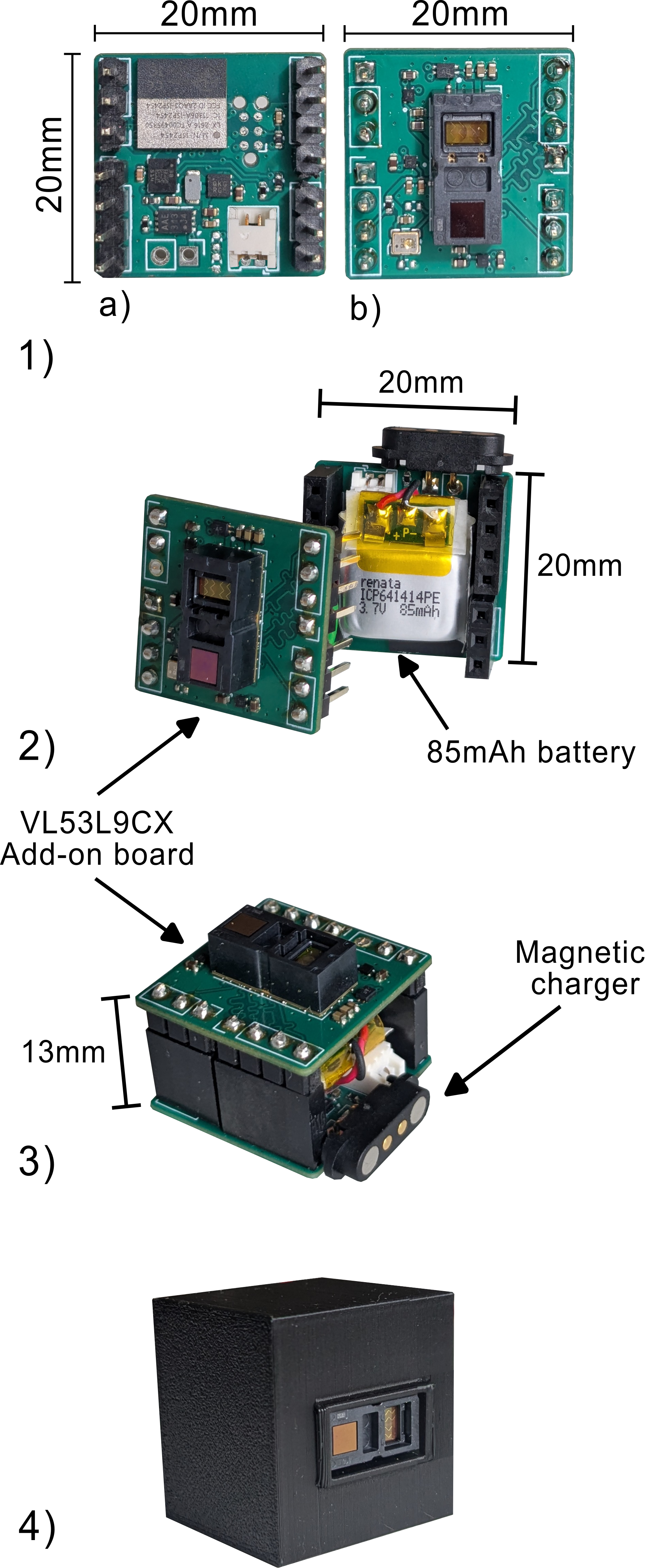}
    \vspace{-1mm}
    \caption{Hardware of the node. 1)~Base board with the \textsc{ISP2454} module (a) and add-on board carrying the \textsc{VL53L9CX} (b), each \qtyproduct{20 x 20}{\milli\meter}. 2)~The two boards stacked with the \qty{85}{\milli\ampere\hour} cell between them. 3)~Assembled stack, \qty{13}{\milli\meter} high, with the magnetic charging connector. 4)~Closed enclosure, whose only opening is the sensor aperture.}
    \label{fig:photos}
    \vspace{-0.1cm}
\end{figure}

\subsection{Spatial Resolution}

Because the node counts people present rather than crossings, the figure of merit is how close two people may stand and still be resolved as two blobs.
With the node at $h=\qty{1.90}{\meter}$, tilted towards
the room, two people are separated down to \qty{8}{\centi\meter} shoulder to shoulder and \qty{60}{\centi\meter} when one stands behind the other. Since F3 merges any edge-adjacent foreground zones regardless of their range, two people are resolved only if at least one background zone separates them in the image.
Laterally, this gap appears as soon as they step apart: at a typical range of about \qty{2.5}{\meter} one zone spans about \qty{44}{\milli\meter}, so the lateral limits are only two to five zones and are set by the sampling grid.

\subsection{Counting Accuracy}
\begin{figure}[t]
    \centering
    \begin{minipage}[b]{\linewidth}
        \centering
        \includegraphics[width=\linewidth]{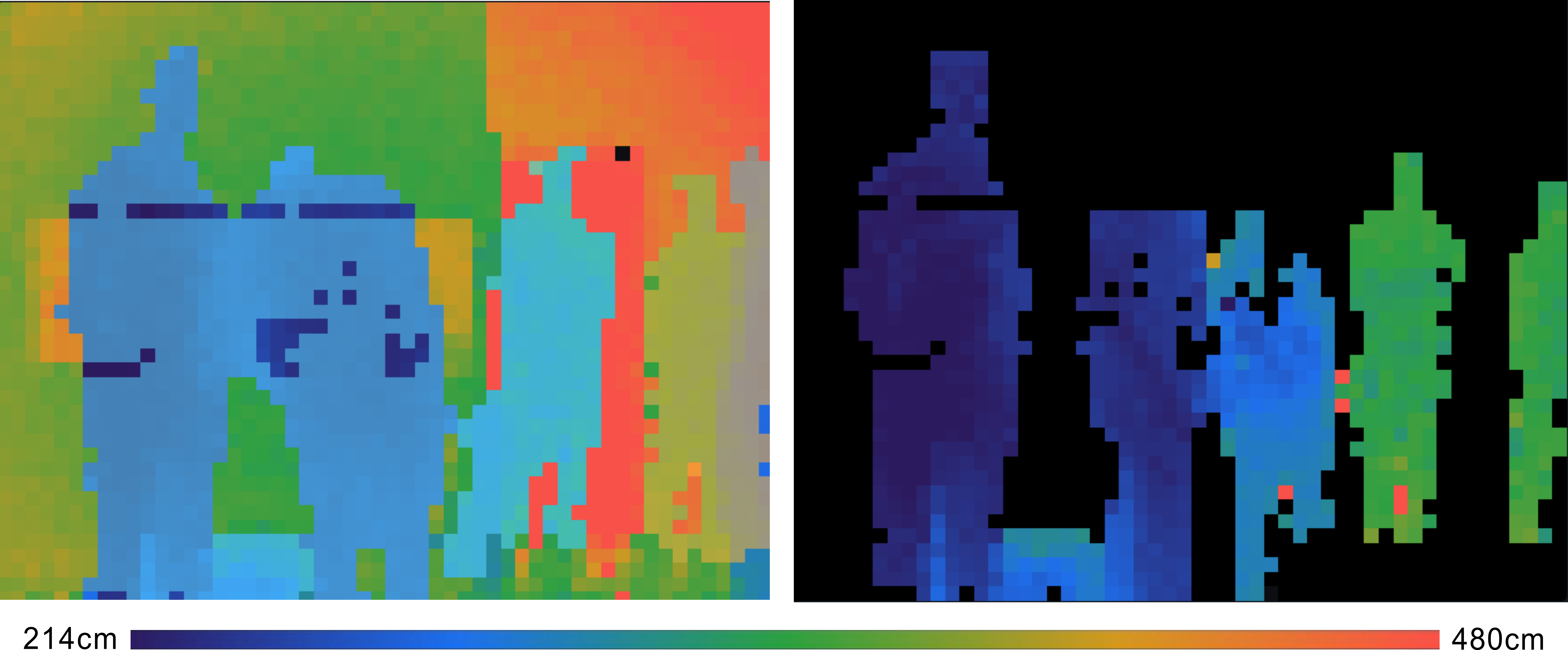}
    \end{minipage}\\[4pt]
        \centering
        \includegraphics[width=\linewidth]{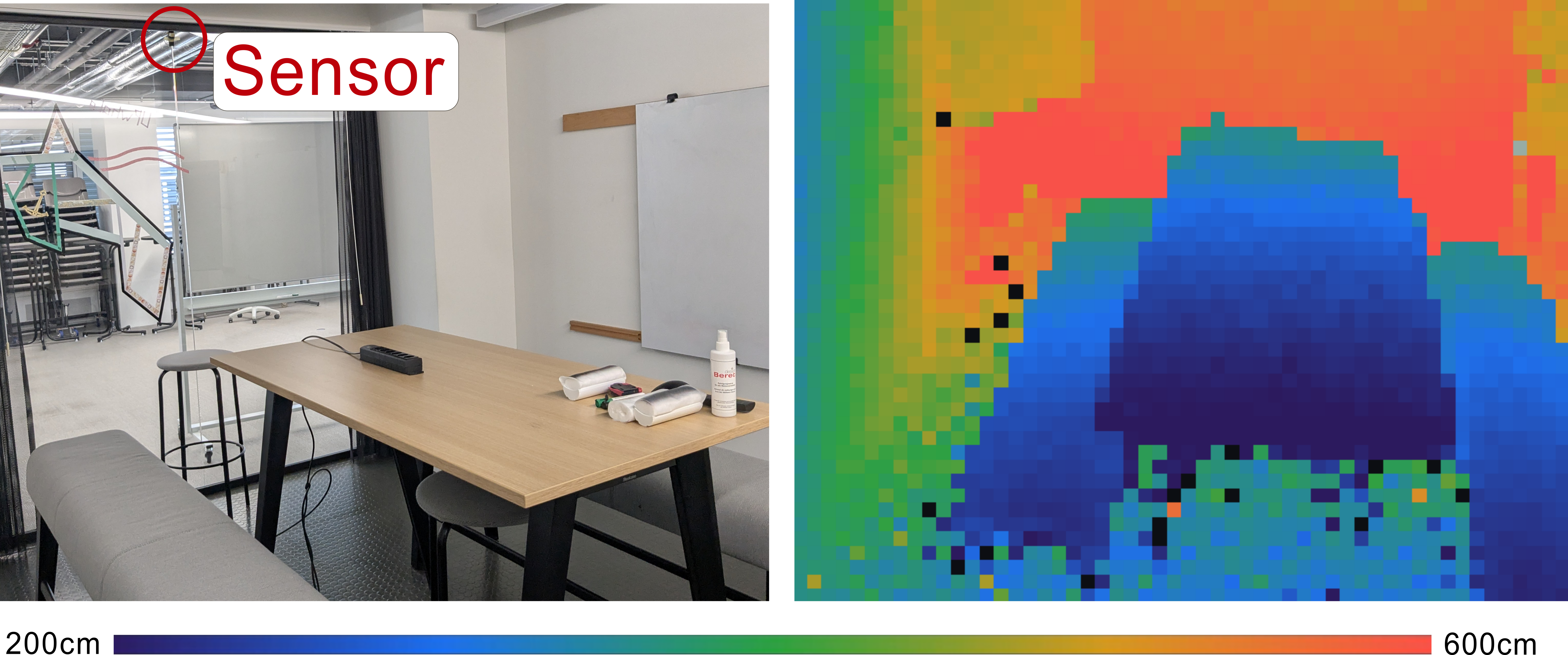}
    \vspace{-3mm}
    \caption{Top: five people side by side in the meeting area, showing their difference in distance to the sensor. Bottom: the meeting-room test setup in which the sensor assesses occupancy.}
    \label{fig:scenes}
    \vspace{-0.5cm}
\end{figure}

Each algorithm's per-frame count is compared with the ground truth using the exact-count accuracy, the share of frames within $\pm1$~person, the \ac{MAE} and the bias. No post-processing filtering is applied. \Cref{fig:scenes} shows a five-person distance frame and the meeting-room setup and \cref{tab:accuracy} gives the results.

\begin{table}[t]
\renewcommand{\arraystretch}{1.1}
\centering
\footnotesize
\caption{Counting accuracy per scenario, on the per-frame count without filtering.}
\label{tab:accuracy}
\vspace{-0.2cm}
\begin{tabular}{@{}lrrr@{}}
\toprule
 & A1 & A2 & A3 \\
\midrule
\multicolumn{4}{@{}l}{\textbf{S1 Meeting room}} \\
\quad Exact count (\unit{\percent}) & 65.7 & 21.3 & 48.7 \\
\quad Within $\pm1$ person (\unit{\percent}) & 99.7 & 39.7 & 79.3 \\
\quad \ac{MAE} (persons) & 0.35 & 2.29 & 0.75 \\
\quad Bias (persons) & $-0.20$ & $-1.92$ & $-0.38$ \\
\midrule
\multicolumn{4}{@{}l}{\textbf{S2 Meeting area, standing}} \\
\quad Exact count (\unit{\percent}) & 62.4 & 17.0 & 59.4 \\
\quad Within $\pm1$ person (\unit{\percent}) & 90.3 & 30.9 & 92.7 \\
\quad \ac{MAE} (persons) & 0.47 & 2.30 & 0.48 \\
\quad Bias (persons) & $-0.36$ & $-1.50$ & $-0.09$ \\
\midrule
\multicolumn{4}{@{}l}{\textbf{S3 Office}} \\
\quad Exact count (\unit{\percent}) & 22.7 & 23.9 & 32.4 \\
\quad Within $\pm1$ person (\unit{\percent}) & 50.0 & 46.0 & 73.3 \\
\quad \ac{MAE} (persons) & 1.52 & 1.93 & 0.98 \\
\quad Bias (persons) & $-1.37$ & $+0.18$ & $-0.52$ \\
\midrule
\multicolumn{4}{@{}l}{\textbf{All scenarios}} \\
\quad Exact count (weighted average) (\unit{\percent}) & 53.0 & 20.9 & 47.0 \\
\quad Within $\pm1$ person (weighted average) (\unit{\percent}) & 83.6 & 39.2 & 81.1 \\
\quad \ac{MAE} (persons) & 0.70 & 2.20 & 0.74 \\
\quad Bias (persons) & $-0.56$ & $-1.23$ & $-0.34$ \\
\bottomrule
\end{tabular}
\vspace{-0.4cm}
\end{table}

\begin{figure}[t]
    \centering
    \centering
    \includegraphics[width=\linewidth]{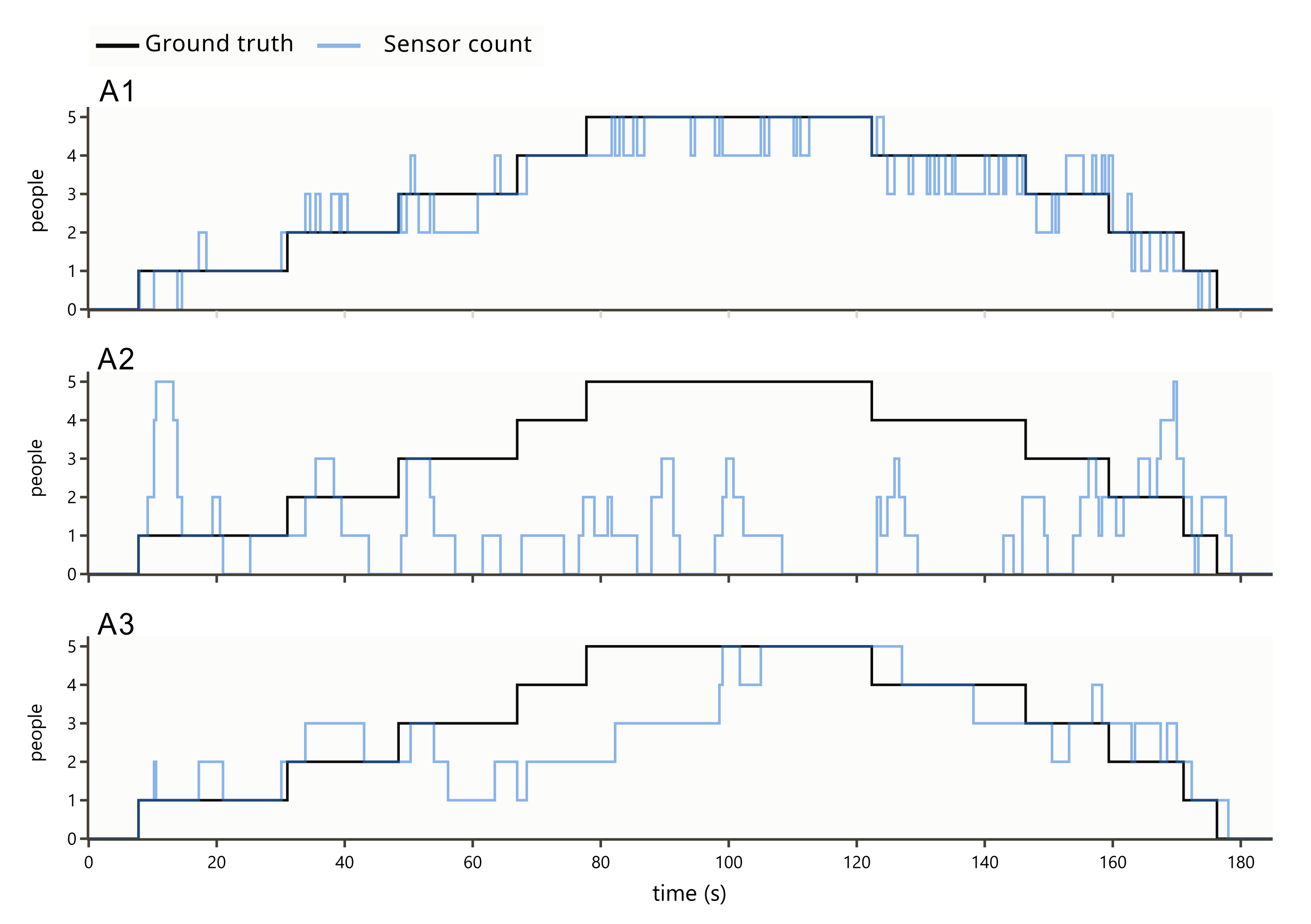}
    \vspace{-8mm}
    \caption{Recorded data of the algorithms A1--A3 in the meeting room as shown in \cref{fig:scenes} bottom.}
    \label{fig:timeline}
    \vspace{-0.2cm}
\end{figure}

In the meeting room (S1), A1 is exact in \qty{65.7}{\percent} of frames and within $\pm1$~person in all but one of the \num{300} recorded frames. In comparison, A3 is exact in \qty{48.7}{\percent} and is within $\pm1$~person in \qty{79.3}{\percent}, and A2 in \qty{21.3}{\percent} and \qty{39.7}{\percent}. The three fail differently (\cref{fig:timeline}): A2 only registers a person while that person moves, so it loses occupants as soon as they settle, as reflected by its bias of $-1.92$~persons, while A1 and A3 keep static occupants. In the standing meeting area (S2), A1 and A3 are close: A1 is exact more often (\qty{62.4}{\percent} against \qty{59.4}{\percent}), while A3 is more often within $\pm1$~person (\qty{92.7}{\percent} against \qty{90.3}{\percent}) and nearly unbiased ($-0.09$~persons).

The office (S3) is the hardest case. Black chairs and large black screens absorb the \qty{940}{\nano\meter} light, so people in front of or on them break into several components, and chairs moved after calibration add further foreground. For instance, A1 never counts more than three people and drops to \qty{22.7}{\percent}, whereas A3, which carries its tracks across frames, is the most accurate here (\qty{32.4}{\percent} exact and \qty{73.3}{\percent} within $\pm1$~person). A2 registers more moving blobs than people and overcounts at low occupancy.

Across all scenarios, A1 performs best with \qty{53.0}{\percent} exact-count accuracy, \qty{83.6}{\percent} of frames within $\pm1$~person and an \ac{MAE} of \num{0.70}~persons. A3 follows with \qty{47.0}{\percent} and the smallest bias, $-0.34$~persons. Both undercount on average in every scenario.

\subsection{Power Consumption}
\label{subsec:power}
\begin{figure}[t]
    \centering
    \includegraphics[width=\linewidth]{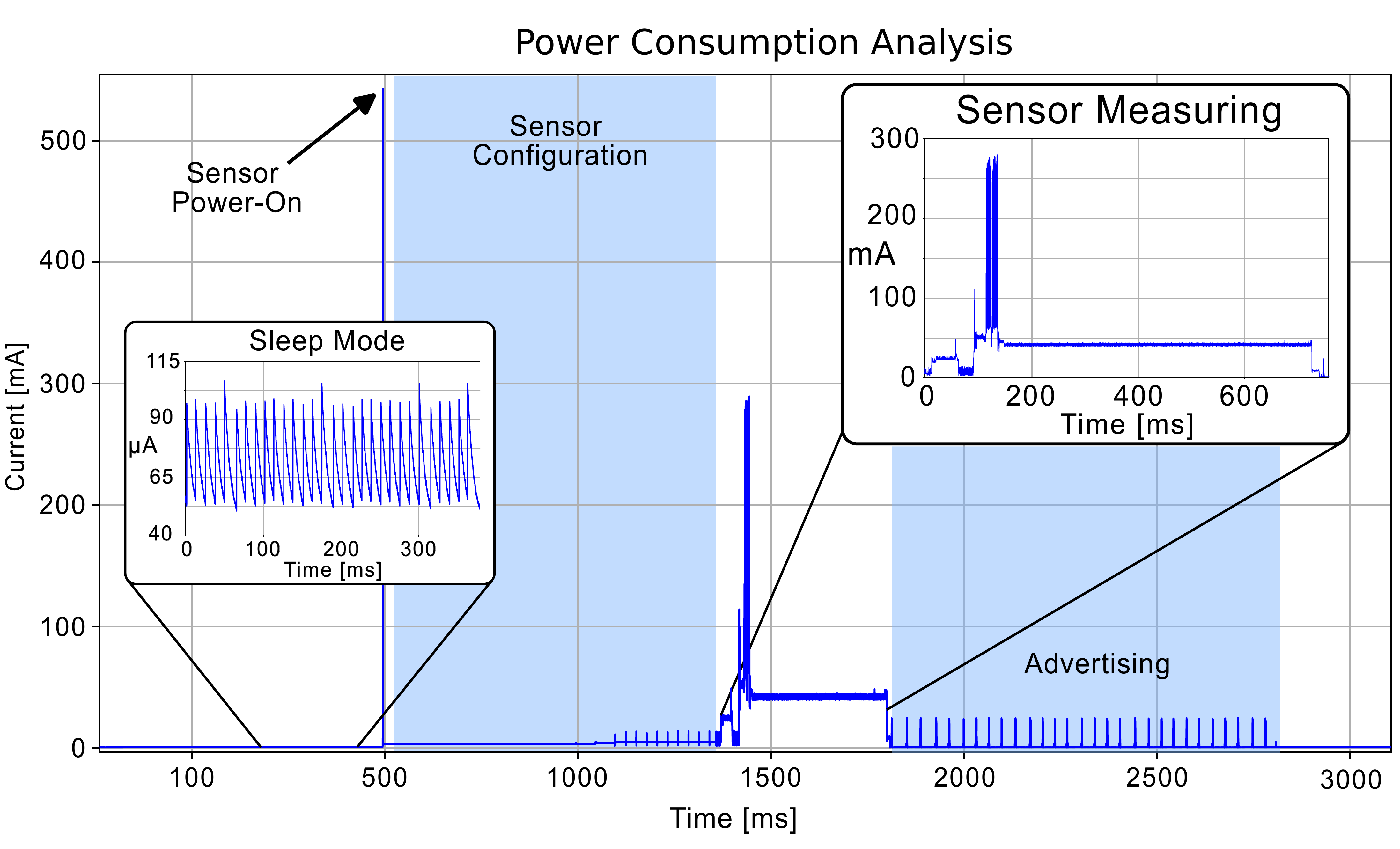}
    \vspace{-5.1mm}
    \caption{One measurement cycle at \qty{3.7}{\volt}. The cycle starts in sleep mode (\qty{282}{\micro\watt}), followed by sensor configuration, the measurement phase (ranging + readout) and data advertisement. The total time for one measurement is \qty{2.31}{\second}. }
    \label{fig:power}
    \vspace{-0.1cm}
\end{figure}

The node was characterised with a \textsc{Power Profiler Kit II} (\textit{Nordic Semiconductor}) at \qty{3.7}{\volt}. \cref{fig:power} shows one measurement cycle and~\cref{tab:power} the required power per state. Average power follows from the charge of one measurement and the interval between them, resulting in a device runtime with a given battery capacity. 

Ranging dominates the power consumption, so the interval is the main lever for low-power operation. At $T=\qty{30}{\second}$ a node running A1 draws \qty{3.1}{\milli\watt} and lasts \qty{60}{\day} on \qty{1200}{\milli\ampere\hour}, against \qty{154}{\milli\watt} and \qty{28.7}{\hour} when set in free-running mode at the maximum frame rate of 2.5 frames per second. The cube of~\cref{fig:photos} with its small form factor has a battery with \qty{85}{\milli\ampere\hour} capacity and a  runtime of \qty{4.2}{\day}, enough for a short survey in a \qtyproduct{20 x 20 x 13}{\milli\meter} package.



Comparing the algorithms, A1 needs no frame history and ranges a single frame ($N=1$), whereas A2 and A3 compare frames and range a burst of $N=5$. Ranging and readout therefore cost \qty{69.8}{\milli\joule} for A1, against \qty{355.4}{\milli\joule} for a burst of $N=5$ frames. Together with the sensor boot, the advertisement and the \qty{282}{\micro\watt} sleep floor, these measurements give the average system power for any duty-cycling interval. At \qty{30}{\second}, A1 draws \qty{3.1}{\milli\watt} and runs \qty{60}{\day}, whereas A2 and A3 draw \qty{12.6}{\milli\watt} and run \qty{15}{\day}.

These figures rely on switching the \textsc{VL53L9CX} off between measurements through its dedicated \acp{LDO}, so the sensor must be rebooted and its firmware patch uploaded again after every wake-up. This boot takes \qty{880}{\milli\second} at \qty{13.9}{\milli\watt}, whereas keeping the sensor powered but idle draws \qty{2.8}{\milli\watt}. Switching the sensor off therefore pays off for measurement intervals longer than about \qty{4.8}{\second}.

\begin{table}[t]
\renewcommand{\arraystretch}{1.0}
\setlength{\tabcolsep}{6pt}
\centering
\footnotesize
\caption{Power evaluation measured at a \qty{3.7}{\volt} supply and the \qty{20}{\milli\second} exposure setting (a frame includes ranging and readout), and the resulting average power and lifetime on a \qty{1200}{\milli\ampere\hour} cell.}
\label{tab:power}
\vspace{-0.2cm}
\begin{tabular}{@{}lrr@{}}
\toprule
State & Duration (\unit{\second}) & Power (\unit{\milli\watt}) \\
\midrule
Sleep, sensor off & --- & 0.282 \\
Sensor boot and configuration & 0.880 & 13.9 \\
One frame & 0.436 & 160.1 \\
Burst of five frames & 1.98 & 179.5 \\
Advertisement & 0.995 & 3.36 \\
Measurement cycle, $N=1$ & 2.31 & 36.9 \\
Measurement cycle, $N=5$ & 3.86 & 96.2 \\
\midrule
Interval $T$ & A1, $N{=}1$ & A2/A3, $N{=}5$ \\
 & \unit{\milli\watt} / \unit{\day} & \unit{\milli\watt} / \unit{\day} \\
\midrule
\qty{5}{\second}  & 17.2 / 10.8 & 74.2 / 2.5 \\
\qty{10}{\second} & 8.7 / 21.2  & 37.3 / 5.0 \\
\qty{30}{\second} & 3.1 / 60    & 12.6 / 15  \\
\midrule
Free running & \multicolumn{2}{r}{\qty{154.6}{\milli\watt} / \qty{28.7}{\hour}} \\
\bottomrule
\end{tabular}
\vspace{-0.4cm}
\end{table}
\section{Conclusion}\label{conclusion}
This paper presented \textit{PepDeTech}, a battery-powered occupancy node that counts people on-chip with a $54\times42$-zone \ac{dToF} \ac{LiDAR} (\textsc{VL53L9CX}) and reports the count through \ac{BLE} advertisements. Unlike existing depth-based counters it needs neither mains power nor a processing computer, and it counts the people present rather than doorway crossings, so errors do not accumulate, and it can provide a real-time occupancy metric in crowded areas. Furthermore, its three training-free algorithms need only a single empty-room calibration of as few as \num{16} frames (about \qty{10}{\second}). Over three scenarios, background subtraction (A1) was the most accurate algorithm, with \qty{53.0}{\percent} exact-count accuracy, \qty{83.6}{\percent} of frames within $\pm1$~person and an \ac{MAE} of \num{0.70}~persons, including static occupants. It is also the most efficient: it ranges one frame per measurement and draws \qty{3.1}{\milli\watt} on average at a \qty{30}{\second} interval. On the \qty{85}{\milli\ampere\hour} cell of the prototype this gives \qty{4.2}{\day}; scaled to a \qty{1200}{\milli\ampere\hour} cell, the measured cycle charge projects to 60 days. Weeks of unattended operation make the node a survey instrument: a manager can characterise one room for a campaign and size lighting, climate, cleaning and space allocation from the record. Its main limit is the field of view, since at a distance of \qty{5}{\meter} the \ac{LiDAR} covers only \qty{5.1}{\meter} $\times$ \qty{3.8}{\meter}, limiting its detection range. In rooms such as offices, large light-absorbing surfaces (black chairs, screens) fragment people and may reduce accuracy. Future work will focus on improving the algorithms and on time-synchronised nodes that extend the field of view by combining multiple sensors.

\bibliographystyle{IEEEtran}
\bibliography{bib/IEEEabrv,bib/references}

@article{Sun_2020,
  author  = {Sun, Kailai and Zhao, Qianchuan and Zou, Jianhong},
  title   = {A review of building occupancy measurement systems},
  journal = {Energy and Buildings},
  volume  = {216},
  pages   = {109965},
  year    = {2020},
  doi     = {10.1016/j.enbuild.2020.109965}
}

@inproceedings{Beltran_2013,
  author    = {Beltran, Alex and Erickson, Varick L. and Cerpa, Alberto E.},
  title     = {{ThermoSense}: Occupancy Thermal Based Sensing for {HVAC} Control},
  booktitle = {Proceedings of the 5th ACM Workshop on Embedded Systems for Energy-Efficient Buildings (BuildSys)},
  pages     = {1--8},
  year      = {2013},
  doi       = {10.1145/2528282.2528301}
}

@inproceedings{Mohammadmoradi_2017,
  author    = {Mohammadmoradi, Hessam and Munir, Sirajum and Gnawali, Omprakash and Shelton, Charles},
  title     = {Measuring People-Flow through Doorways Using Easy-to-Install {IR} Array Sensors},
  booktitle = {2017 13th International Conference on Distributed Computing in Sensor Systems (DCOSS)},
  pages     = {35--43},
  year      = {2017},
  doi       = {10.1109/DCOSS.2017.26}
}

@inproceedings{Munir_2017,
  author    = {Munir, Sirajum and Arora, Ripudaman Singh and Hesling, Craig and Li, Juncheng and Francis, Jonathan and Shelton, Charles and Martin, Christopher and Rowe, Anthony and Berges, Mario},
  title     = {Real-Time Fine Grained Occupancy Estimation Using Depth Sensors on {ARM} Embedded Platforms},
  booktitle = {2017 IEEE Real-Time and Embedded Technology and Applications Symposium (RTAS)},
  pages     = {295--306},
  year      = {2017},
  doi       = {10.1109/RTAS.2017.8}
}

@article{Lu_2021,
  author  = {Lu, Hao and Tuzikas, Arunas and Radke, Richard J.},
  title   = {A zone-level occupancy counting system for commercial office spaces using low-resolution time-of-flight sensors},
  journal = {Energy and Buildings},
  volume  = {252},
  pages   = {111390},
  year    = {2021},
  doi     = {10.1016/j.enbuild.2021.111390}
}

@article{Xie_2023,
  author  = {Xie, Chen and Daghero, Francesco and Chen, Yukai and Castellano, Marco and Gandolfi, Luca and Calimera, Andrea and Macii, Enrico and Poncino, Massimo and Jahier Pagliari, Daniele},
  title   = {Efficient Deep Learning Models for Privacy-Preserving People Counting on Low-Resolution Infrared Arrays},
  journal = {IEEE Internet of Things Journal},
  volume  = {10},
  number  = {15},
  pages   = {13895--13907},
  year    = {2023},
  doi     = {10.1109/JIOT.2023.3263290}
}

@article{Jeong_2025,
  author  = {Jeong, Jaeik and Park, Wan-Ki},
  title   = {Privacy-preserving labeling-free occupancy counting sensor based on {ToF} camera and clustering},
  journal = {ETRI Journal},
  volume  = {47},
  number  = {5},
  pages   = {841--854},
  year    = {2025},
  doi     = {10.4218/etrij.2025-0022}
}

@article{Huang_2021,
  author  = {Huang, Xu and Cheena, Hasnain and Thomas, Abin and Tsoi, Joseph K. P.},
  title   = {Indoor Detection and Tracking of People Using {mmWave} Sensor},
  journal = {Journal of Sensors},
  volume  = {2021},
  pages   = {6657709},
  year    = {2021},
  doi     = {10.1155/2021/6657709}
}

@misc{ST_VL53L9CX,
  author       = {{STMicroelectronics}},
  title        = {{VL53L9CX}: Direct Time-of-Flight {3D} {LiDAR} Module with up to 2.3k Zones},
  howpublished = {Product page and datasheet},
  url          = {https://www.st.com/en/imaging-and-photonics-solutions/vl53l9cx.html},
  note         = {Accessed: 2026-10-03},
  year         = {2026}
}

@manual{ST_UM3683,
  author       = {{STMicroelectronics}},
  title        = {{UM3683}: {VL53L9CX} Programming Guide, Rev. 3},
  organization = {STMicroelectronics},
  year         = {2026},
  note         = {{Table}~23: recommended configuration}
}

@misc{Nordic_nRF54L15,
  author       = {{Nordic Semiconductor}},
  title        = {{nRF54L15} Wireless {SoC}},
  howpublished = {Product page},
  url          = {https://www.nordicsemi.com/Products/nRF54L15},
  note         = {Accessed: 2026-10-03},
  year         = {2026}
}

@misc{InsightSiP_ISP2454,
  author       = {{Insight SiP}},
  title        = {{ISP2454} {Bluetooth Low Energy} Module Datasheet},
  howpublished = {Datasheet},
  url          = {https://www.insightsip.com/products/bluetooth-le-modules/isp2454},
  note         = {Accessed: 2026-10-03},
  year         = {2025}
}

@misc{Luder_2025,
  author        = {Luder, Victor and Magno, Michele},
  title         = {Self-Sustaining Multi-Sensor {LoRa}-Based Activity Monitoring for Community Workout Parks},
  year          = {2025},
  eprint        = {2506.03203},
  archivePrefix = {arXiv},
  primaryClass  = {cs.NI},
  url           = {https://arxiv.org/abs/2506.03203}
}

@inproceedings{Raykov_2016,
  author    = {Raykov, Yordan P. and Ozer, Emre and Dasika, Ganesh and Boukouvalas, Alexis and Little, Max A.},
  title     = {Predicting Room Occupancy with a Single Passive Infrared ({PIR}) Sensor through Behavior Extraction},
  booktitle = {Proceedings of the 2016 ACM International Joint Conference on Pervasive and Ubiquitous Computing (UbiComp)},
  pages     = {1016--1027},
  year      = {2016},
  doi       = {10.1145/2971648.2971746}
}

@article{Depatla_2015,
  author  = {Depatla, Saandeep and Muralidharan, Arjun and Mostofi, Yasamin},
  title   = {Occupancy Estimation Using Only {WiFi} Power Measurements},
  journal = {IEEE Journal on Selected Areas in Communications},
  volume  = {33},
  number  = {7},
  pages   = {1381--1393},
  year    = {2015},
  doi     = {10.1109/JSAC.2015.2430272}
}

@article{Mikkilineni_2019,
  author  = {Mikkilineni, Aravind K. and Dong, Jin and Kuruganti, Teja and Fugate, David},
  title   = {A novel occupancy detection solution using low-power {IR-FPA} based wireless occupancy sensor},
  journal = {Energy and Buildings},
  volume  = {192},
  pages   = {63--74},
  year    = {2019},
  doi     = {10.1016/j.enbuild.2019.03.022}
}

@inproceedings{Bhattacharya_2016,
  author    = {Bhattacharya, Indrani and Radke, Richard J.},
  title     = {Arrays of single pixel time-of-flight sensors for privacy preserving tracking and coarse pose estimation},
  booktitle = {2016 IEEE Winter Conference on Applications of Computer Vision (WACV)},
  pages     = {1--9},
  year      = {2016},
  doi       = {10.1109/WACV.2016.7477671}
}

@inproceedings{Piccardi_2004,
  author    = {Piccardi, Massimo},
  title     = {Background subtraction techniques: a review},
  booktitle = {2004 IEEE International Conference on Systems, Man and Cybernetics},
  pages     = {3099--3104},
  year      = {2004},
  doi       = {10.1109/ICSMC.2004.1400815}
}

@article{Bobick_2001,
  author  = {Bobick, Aaron F. and Davis, James W.},
  title   = {The recognition of human movement using temporal templates},
  journal = {IEEE Transactions on Pattern Analysis and Machine Intelligence},
  volume  = {23},
  number  = {3},
  pages   = {257--267},
  year    = {2001},
  doi     = {10.1109/34.910878}
}

@article{Wu_2009,
  author  = {Wu, Kesheng and Otoo, Ekow and Suzuki, Kenji},
  title   = {Optimizing two-pass connected-component labeling algorithms},
  journal = {Pattern Analysis and Applications},
  volume  = {12},
  number  = {2},
  pages   = {117--135},
  year    = {2009},
  doi     = {10.1007/s10044-008-0109-y}
}

@inproceedings{Stauffer_1999,
  author={Stauffer, C. and Grimson, W.E.L.},
  booktitle={Proceedings. 1999 IEEE Computer Society Conference on Computer Vision and Pattern Recognition (Cat. No PR00149)}, 
  title={Adaptive background mixture models for real-time tracking}, 
  year={1999},
  volume={2},
  number={},
  pages={246-252},
  doi={10.1109/CVPR.1999.784637}
}

@inproceedings{Bewley_2016,
 author={Bewley, Alex and Ge, Zongyuan and Ott, Lionel and Ramos, Fabio and Upcroft, Ben},
  booktitle={2016 IEEE International Conference on Image Processing (ICIP)}, 
  title={Simple online and realtime tracking}, 
  year={2016},
  volume={},
  number={},
  pages={3464-3468},
  doi={10.1109/ICIP.2016.7533003}
  }

\end{document}